\documentclass[prb,twocolumn,aps,floats,floatfix,superscriptaddress]{revtex4}
\usepackage{graphicx,epsfig}

\begin{document}

\title{Icosadeltahedral geometry of fullerenes, viruses and geodesic domes}
\author{Antonio \v{S}iber}
\email{asiber@ifs.hr}
\affiliation{Department of Theoretical Physics, Jo\v{z}ef Stefan Institute, SI-1000 Ljubljana, Slovenia}
\affiliation{Institute of Physics, P.O. Box 304, 10001 Zagreb, Croatia}

\begin{abstract}
I discuss the symmetry of fullerenes, viruses and geodesic domes within a unified framework of 
icosadeltahedral representation of these objects. The icosadeltahedral symmetry is explained in details by 
examination of all of these structures. Using Euler's theorem on polyhedra, it is shown 
how to calculate the number of vertices, edges, and faces in domes, and number 
of atoms, bonds and pentagonal and hexagonal rings in fullerenes. Caspar-Klug 
classification of viruses is elaborated as a specific case of icosadeltahedral geometry.
\end{abstract}
\maketitle

\section{Introduction}

Although the fullerene molecules were previously predicted and discussed \cite{Krotolecture}, it was quite a surprise 
for largest part of the scientific community when Kroto {\em et al} published a paper announcing their experimental 
discovery \cite{discoveryfull}. Some part of this surprise can surely be attributed to amazingly symmetrical 
arrangement of carbon atoms in the most abundant of all the fullerene molecules - Buckminsterfullerene or C$_{60}$. 
Kroto and colleagues named this molecule after Richard Buckminster Fuller who was an 
American designer, inventor and architect\cite{discoveryfull} most recognized for popularizing  
geodesic domes - networks of interconnected struts forming a (hemi)spherical grid. Fuller constructed 
domes as an alternative to architecture that was dominant in 1960's in USA. Most famous of them was used 
to house American pavilion for the World Expo exhibition in Montreal in year 1967. Geodesic domes that Fuller 
constructed were not chiral \cite{casparfullerenes}, i.e. their mirror image retained their symmetry. 
Even before the discovery of fullerenes it became clear that achiral domes that Fuller constructed belong 
to a larger class of mathematical structures called icosadeltahedra. Extending Fuller's 
design ideas\cite{Morgan}, in year 1962 Donald Caspar and Aaron Klug constructed the first theory that explained 
the features of most of the so-called ''spherical'' (icosahedral) viruses known at the time \cite{casparklug}. 

It is intriguing that objects so different in size (fullerenes ~ 1 nm, viruses ~ 100 nm, and 
geodesic domes ~ 100 m) share the same design and symmetry. The reason for this is that all 
these objects are built up of nearly identical elements (carbon atoms in fullerenes, 
proteins in viruses, and struts in geodesic domes) that have to arrange so 
to fully enclose the object interior i.e. to form a cage-like structure. The aim of this 
paper is to explain the 
design principles and symmetries of a large subset of these structures that has an 
icosadeltahedral symmetry.

\section{Icosadeltahedral geodesic domes}

\begin{figure*}[ht]
\centerline{
\epsfig {file=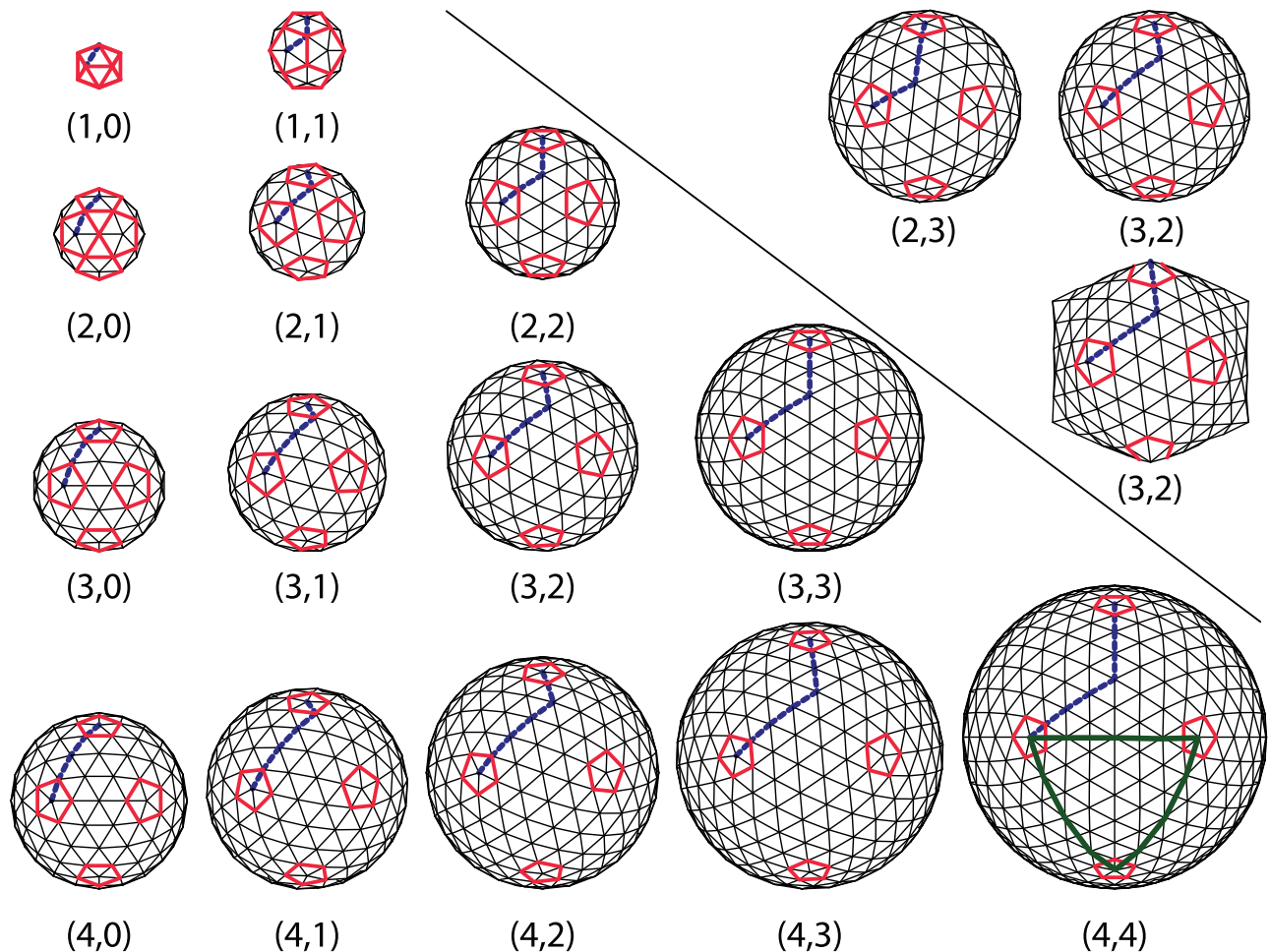,width=17cm}
}
\caption{Gallery of icosadeltahedral geodesic domes for $m>n$ and $m<5$. For the sake of clarity, the back sides of geodesic domes are not shown, i.e. 
only half of the dome and four of twelve icosahedral vertices can bee seen. The upper-right corner of the figure 
contains comparison between chiral $(3,2)$ and $(2,3)$ domes. Note that they are mirror images. The spiky shape 
is a $(3,2)$ icosadeltahedron (all of its faces are equilateral triangles).} 
\label{fig:fig1}
\end{figure*}

Icosadeltahedral geodesic domes 
can be mathematically described as triangulations of the spherical surface (or a part of it) with the 
icosahedral ''backbone''. A visual representation of this statement is shown in Fig. \ref{fig:fig1}. Icosahedron 
(shown in the upper left corner of Fig. \ref{fig:fig1}) is a platonic solid consisting of twenty equilateral triangles 
and twelve vertices. When the spherical surface is covered with triangles so that the icosahedral nature of 
the triangulation is preserved, the twelve icosahedral vertices become special points. These twelve 
points become the only ones that have {\em five} nearest neighboring points - all the other points have {\em six} 
nearest neighbors. In Fig. \ref{fig:fig1} the neighboring points of the icosahedral vertices are outlined as 
thick pentagons. As Fig. \ref{fig:fig1} illustrates, there are many ways to triangulate a sphere so that there 
are twelve points that make the vertices of an icosahedron and have five nearest neighbors, all other points having 
six nearest neighboring points. All of these 
triangulations are called {\em icosadeltahedral}. Deltahedron is a polyhedron whose faces are all equilateral 
triangles. Strictly speaking the icosadeltahedral geodesic domes are {\em not} (icosa)deltahedra since all of 
the triangles that they consist of are not equilateral. The requirement that all of the polyhedral faces 
be equilateral triangles necessarily produces aspherical (spiky) polyhedra. 
Nevertheless, the icosadeltahedral geodesic domes have the same symmetry as an icosadeltahedron 
which is a spiky shape shown in Fig. \ref{fig:fig1}. In fact the domes can be considered as projections of 
icosadeltahedra on their circumscribed spheres. In the following the word dome is used only for 
icosadeltahedral geodesic dome.

Each of the domes can be characterized by two nonnegative integers, denoted by $m$ and $n$ in the following. These can be thought 
of as numbers of ''jumps'' through the vertices of a dome that need to be performed in order to reach 
a center of a pentagon from its neighboring pentagon. Except for $(m,0)$ dome, the jumps need to be directed 
along two different spherical geodesics (the shortest lines between two points on a sphere), $m$ along one of them, 
and $n$ along another one, making an angle of 60 degrees with the first one. To 
be definite, we need to specify whether the ''jumper'' needs to turn left or right after the $m$ jumps along the 
first spherical geodesic. In what follows, I shall assume the left turn and denote the symmetry of 
icosadeltahedral structures by $(m,n)$. Were the other convention chosen, the $(m,n)$ dome in 
our convention would correspond to $(n,m)$ dome in the alternative convention. The 
domes with $m \ne n$; $m,n > 0$ are {\em chiral}. This means that their mirror image has different 
symmetry. A mirror image of $(m,n)$ dome is $(n,m)$ dome. This is illustrated in the upper-right corner of 
Fig. \ref{fig:fig1} for $(3,2)$ dome. 

From $m$ and $n$ one can calculate the number of triangles in a dome. 
The number of triangles per one 
spherical segment bounded by three spherical geodesic passing through neighboring 
fivefold coordinated vertices (outlined on a $(4,4)$ dome in Fig. \ref{fig:fig1}) is 
\begin{equation}
T = m^2 + n^2 + mn,
\label{eq:tnumber}
\end{equation}
so that the total number of triangles (or faces) in an icosadeltahedron is 
\begin{equation}
f = 20 T.
\end{equation}
$T$ is called the triangulation number or simply the T-number. It adopts special integer 
values, $T=1, 3, 4, 7, 9, 12, 13, ...$. Instead of $m$ and $n$ integers, the T-number can be used to classify 
the icosadeltahedral symmetry. The problem with this choice is that it doesn't discriminate between 
$(m,n)$ and $(n,m)$ domes. That is why the T-number is sometimes used in combination with words {\em laevo} (left) and 
{\em dextro} (right) to resolve this ambiguity. For example, $(2,1)$ structure in our convention would 
be in this case denoted as $T=7_{laevo}$ or $T=7_l$ or simply $T=7$, while $(1,2)$ structure 
would be denoted as $T=7_{dextro}$ or $T=7_{d}$ \cite{Bakerreview}. From the known number of polyhedron 
faces ($f$) one can proceed to find the number 
of its vertices ($v$) and edges ($e$) by using Euler's theorem on polyhedra \cite{Euler} which relates these 
nonzero integer quantities as
\begin{equation}
v - e + f = 2.
\label{eq:Euler}
\end{equation}
This equation is valid for polyhedra that are homeomorphic to the sphere, i.e. their topology is 
the same as that of a sphere, which is the case of interest to us. In icosadeltahedral domes twelve 
vertices belong to five edges - these are located at the vertices of an icosahedron. 
All the other vertices belong to six edges, i.e. six edges meet at those vertices. Each edge is bounded 
by two vertices and all these fact together can be used to relate $e$ and $v$ as 
\begin{equation}
2 e = 5\cdot 12 + 6  \cdot(v-12) = 60 + 6(v-12).
\end{equation}
In combination with Euler's theorem, one obtains that 
\begin{equation}
v = \frac{f}{2} + 2 = 10T + 2,
\end{equation}
and 
\begin{equation}
e = \frac{3 f}{2} = 30 T.
\end{equation}

\section{Icosahedral fullerenes}

\begin{figure*}[ht]
\centerline{
\epsfig {file=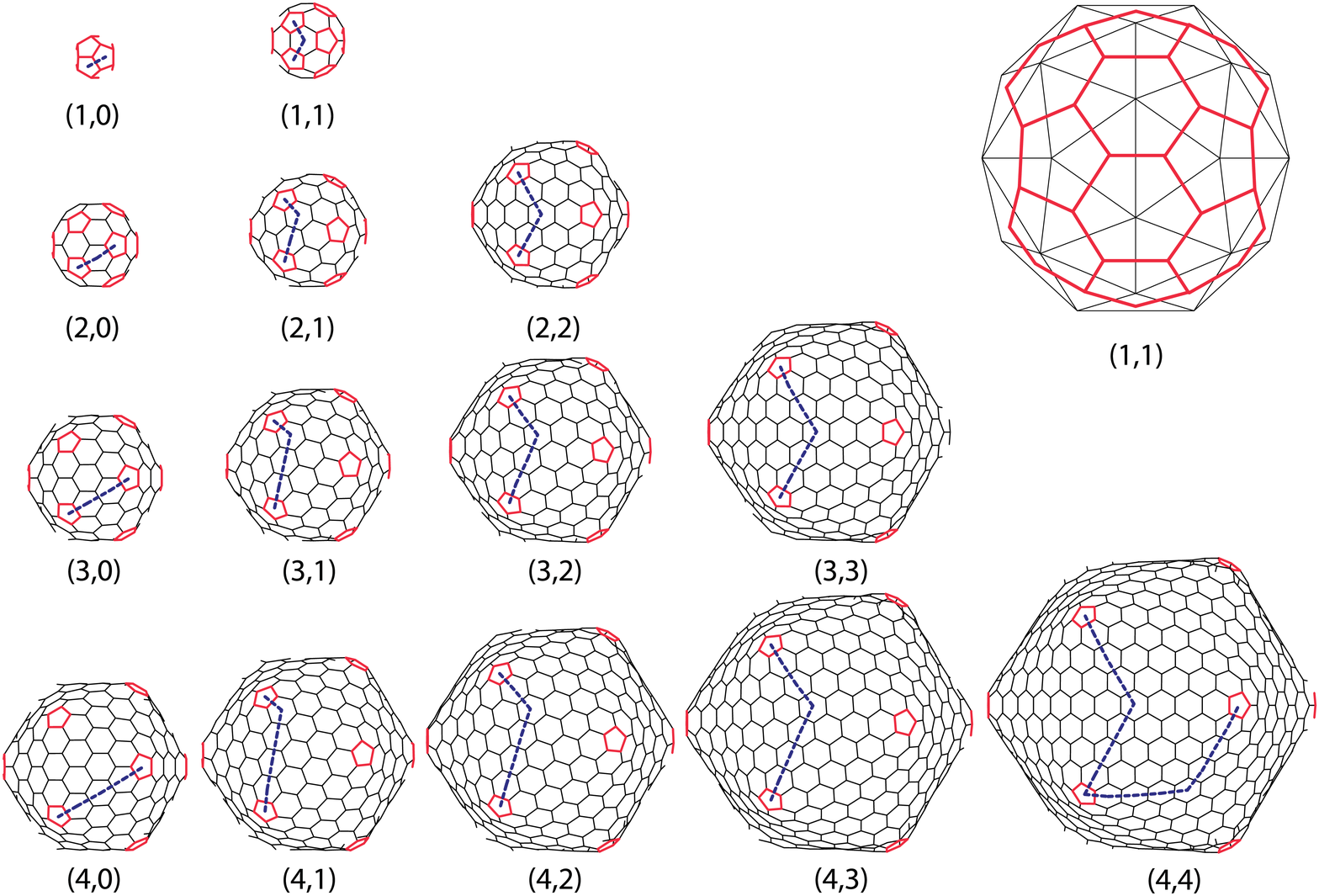,width=17cm}
}
\caption{Gallery of icosahedral fullerenes for $m>n$ and $m<5$. For the sake of clearer representation the back sides of fullerenes are not shown. 
The carbon-carbon bond in all fullerenes is practically everywhere equal to 1.42 \AA, and this can be used to estimate their 
size. The upper-right corner of the figure 
contains comparison between $(1,1)$ dome and $(1,1)$ icosahedral fullerene (Buckminsterfullerene, not to scale with other depicted 
fullerenes). Note that these polyhedra are dual to each other.} 
\label{fig:fig2}
\end{figure*}

Fullerene molecules are carbon cages in which all carbon rings are either pentagonal or hexagonal 
and all carbon atoms make three covalent bonds with their nearest neighbors. This type of bonding 
is also present in the planes of carbon atoms in graphite (graphene planes). It is called sp$^2$ bonding, in contrast 
to sp$^3$ bonding that is characteristic of diamond (in diamond each of the carbon atoms is bonded with {\em three} 
nearest neighboring atoms). There are many different structures that can be made of carbon atoms connected with 
sp$^2$ bonds, at least conceptually (see e.g. Refs. \onlinecite{Sibnano1,Sibnano2}). Quite a 
different question is whether such structures can be experimentally 
obtained. The icosadeltahedral symmetry of geodesic domes is characteristic of a class of 
especially symmetric fullerene molecules, sometimes called icosahedral fullerenes, or giant icosahedral fullerenes 
in case that molecules contain more than about 100 carbon atoms. Buckminsterfullerene belongs to this class 
(its ''companion'' molecule C$_{70}$ that was discovered simultaneously \cite{discoveryfull} does not, however).
The symmetry of these carbon molecules can be obtained from icosadeltahedral domes by placing carbon atoms 
in (bary)centers of every triangle in the dome. The newly obtained set of points (carbon atoms) is then ordered so 
that each point is connected with its three nearest neighbors, i.e. the carbon-carbon bonds are established (this 
procedure is illustrated in the upper-right corner of Fig. \ref{fig:fig2}). 
This fulfills the basic chemical requirement for carbon atoms in sp$^2$ bonding electronic configuration. The thus 
obtained structure contains now twelve pentagonal carbon rings (pentagons) and certain number of hexagonal carbon rings 
(hexagons), depending on the T-number of the dome. The starting dome and the fullerene-like 
polyhedron that was obtained from it are called {\em dual} polyhedra. The 
$(m,n)$ symmetry that was characteristic of the dome will also be 
characteristic of its dual fullerene-like polyhedron, but now the ''jumping'' that characterizes icosadeltahedral 
symmetry is allowed only through the centers of pentagons and hexagons (not along the carbon-carbon bonds). 
I have used the term ''fullerene-like polyhedron'' since the true 
fullerene molecules will in general be different from the polyhedron obtained by a simple mathematical dualization of the dome. 
The domes are icosadeltahedral triangulations of the sphere and this necessarily means that the lengths of triangular 
edges are not all the same. In particular, they are considerably shorter in the neighborhood of icosahedral vertices, and 
this will pertain also in the dual polyhedra. This feature is also characteristic of 
Fuller's geodesic constructions \cite{discoveryfull}. However, the fullerene molecules are more than mathematical 
entities and their exact shape is determined by energetics of carbon-carbon interactions. Carbon-carbon 
bonds are much easier to bend than to stretch \cite{Sibnano1}, so the shape of the fullerene molecule will be such to keep the 
nearest-neighbor carbon-carbon distances as uniform as possible and as close to their equilibrium value as possible 
(the equilibrium length of carbon-carbon bonds in infinitely large graphene plane is about 1.42 \AA). This means that large enough 
fullerenes will necessarily be aspherical, looking more like an icosahedron with vertices slightly above the centers of 
carbon pentagons as the molecules get larger \cite{primjedba1}. Figure \ref{fig:fig2} displays a gallery of icosahedral fullerenes. 
Their shape is not merely a mathematical construction obtained by dualization of a dome, but a 
true minimum of energy, calculated by using the realistic model of energetics of carbon sp$^2$ bonding \cite{Brenner} as 
described in Ref. \onlinecite{Sibnano1}. Note that the Buckminsterfullerene $(1,1)$ is perfectly spherical, i.e. all of 
its carbon atoms are equally distanced from the geometrical center of the molecule. A high degree of sphericity is 
also present in $(2,0)$ fullerene, but already in $(2,1)$ fullerene a clear icosahedral shape of the molecule 
develops and this becomes more prominent in larger molecules. There is a long standing debate (perhaps of 
academic value only) concerning the shape of {\em asymptotic} icosahedral fullerenes, i.e. those that contain 
extremely large (infinite) number of carbon atoms. Studies based on continuum elasticity of the icosadeltahedral 
shells \cite{Witten,onadruga} and on microsopic models of carbon-carbon bonding \cite{SibEPJ} predict that the 
asymptotic shape is a perfect icosahedron.

\begin{figure}[ht]
\centerline{
\epsfig {file=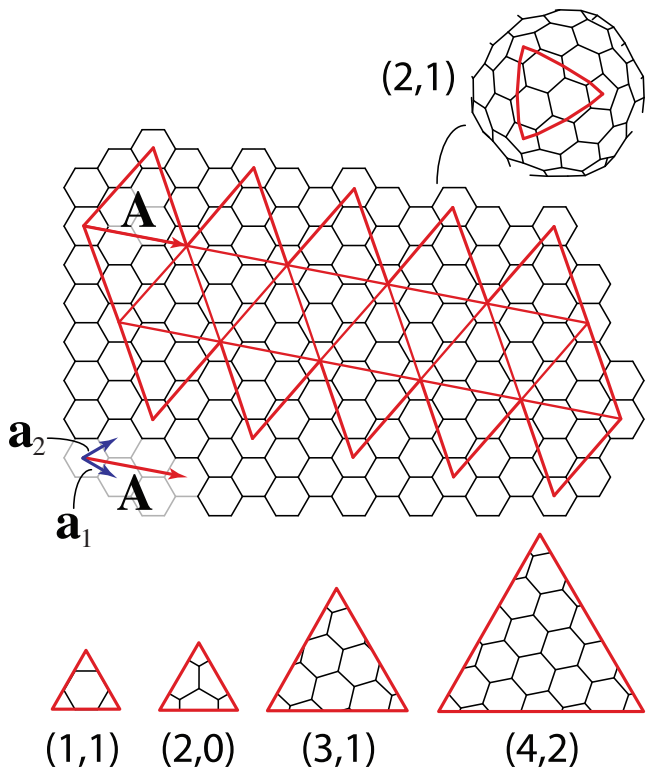,width=8.5cm}
}
\caption{Cut-and-fold construction of $(2,1)$ icosahedral fullerene. The vectors ${\bf a}_1$, 
${\bf a}_2$ and ${\bf A}$ discussed in the text are denoted. The triangular faces of 
$(1,1)$, $(2,0)$, $(3,1)$, and $(4,2)$ fullerene-like icosahedra are shown in the bottom of the figure.} 
\label{fig:fig3}
\end{figure}

A way to better comprehend the symmetry of fullerenes is to ''unfold'' them so that they become polygonal 
pieces of graphene. Alternatively, one can also think about this procedure, illustrated in Fig. \ref{fig:fig3} 
as a way to construct these molecules. The concave polygonal shape consisting of 20 equilateral triangles outlined by thick lines 
is cut out from the graphene plane. The polygon is then creased along the edges shared by the triangles and 
folded into a perfect icosahedron. The thus obtained shape is still not a 
fullerene since the details of its shape are wrong, but it has the same icosadeltahedral symmetry and the same 
connectivity and number of carbon atoms as the icosahedral fullerene does. The integers $m$ and $n$ that 
characterize the shape can now be interpreted as components of a two-dimensional vector ${\bf A}$ in a basis of graphene unit 
cell vectors ${\bf a}_1$ and ${\bf a}_2$ denoted in Fig. \ref{fig:fig3},
\begin{equation}
{\bf A} = m {\bf a}_1 + n {\bf a}_2, \; m,n >0.
\end{equation}
The ${\bf A}$ vector is directed 
along the side of one of the twenty triangles making the icosahedron as illustrated in Fig. \ref{fig:fig3}. In 
this convention, the unit cell vectors ${\bf a}_1$ and ${\bf a}_2$ need to be chosen so that their vector 
product points from the paper towards the reader. This reproduces the jumping-to-the-left convention 
discussed in the previous section. The unfolding described here can also be applied to icosadeltahedra. 
The underlying lattice is triangular in that case. This procedure is very convenient for counting 
faces, edges and vertices and can be used to derive Eq. (\ref{eq:tnumber}).

An important piece of information on fullerene molecules can be obtained from the Euler's theorem on polyhedra. 
Since exactly three bonds (or polyhedron edges) finish at each of the carbon atoms (polyhedron vertices) and the 
bond (edge) is shared by two atoms (vertices), it follows that
\begin{equation}
2e = 3v.
\end{equation}
By definition, the fullerenes contain only pentagonal and hexagonal faces (carbon rings). Let us denote the 
number of pentagonal and hexagonal faces by $f_5$ and $f_6$, respectively. The total number of faces is 
obviously given by
\begin{equation}
f = f_5 + f_6.
\end{equation}
Pentagonal and hexagonal faces are bounded by five and six vertices (atoms), respectively and each vertex (atom) 
belongs to exactly three faces. This means that 
\begin{equation}
5f_5 + 6f_6 = 3v.
\end{equation}
Combining these equations with Euler's theorem in Eq.(\ref{eq:Euler}) one obtains that 
\begin{equation}
f_5 = 12.
\end{equation}
This is obviously true for the icosahedral fullerenes discussed so far, but the equation holds for 
general fullerenes as long as they are topologically equivalent to a sphere (including e.g. C$_{70}$). In 
other words, every network of pentagonal and hexagonal ring of carbon atoms with spherical topology 
necessarily has five pentagonal rings. A relation between the number of carbon atoms in fullerenes 
and the number of hexagonal faces can also be obtained from the above consideration. It states that 
\begin{equation}
v = 2(f_6 + 10).
\end{equation}
This means that the number of carbon atoms in the fullerene molecules is necessarily {\em even}. 
This, at first puzzling, piece of information was observed already in the mass spectra of carbon 
clusters obtained by laser vaporization from the graphitic sample \cite{Curllecture}. Only signatures 
of clusters containing even number of carbon atoms were detected which can be nicely explained 
by assuming that the clusters detected were in fact fullerenes. Let us now specify the discussion 
of general fullerenes to the case of icosahedral fullerenes. 
Total number of carbon atoms in these molecules is 
\begin{equation}
v = 20 (m^2 + mn + n^2) = 20 T,
\end{equation}
and the number of carbon-carbon bonds is 
\begin{equation}
e = 30 T.
\end{equation}
As shown earlier, there are exactly twelve pentagonal carbon rings and 
\begin{equation}
f_6 = 10 (T-1)
\end{equation}
hexagonal carbon rings.

Fullerenes with number of carbon atoms larger than about 120 were not clearly observed in 
the experiments described in Ref. \onlinecite{discoveryfull}. Thus, in addition to C$_{60}$, 
only the C$_{80}$ signature may in fact correspond to ($m=2$, $n=0$) icosahedral fullerene. Larger (giant) 
fullerenes can be observed in experiments but it is difficult to precisely determine their geometry and 
the spatial distribution of pentagonal carbon rings \cite{recentprlfull}. Carbon pentagons 
in graphitic samples have been observed using scanning tunneling microscopy \cite{APLpentagon}.

\section{Caspar-Klug clasification of viruses: The T-number}

Viruses are particles made of DNA or RNA molecule (genome) protected by a coating made of proteins. Their 
size depends on the type of a virus \cite{Bakerreview}. The typical diameter of a virus is about 50 nm. For example the 
diameter of a herpes simplex virus is 125 nm, while polio virus is only 32 nmm in diameter \cite{Bakerreview}. 
The information 
that is required to produce the proteins of the coating is contained in the viral RNA or DNA 
molecule. Once the virus penetrates the cell wall, the viral genome is delivered to the cell and the 
production of viral proteins starts. The proteins are produced by a cellular molecular machinery called 
ribosomes that can assemble proteins from amino acids by ''reading'' the information on the 
required sequence of amino acids that is coded in the viral RNA molecule. This process is called 
{\em translation} by biologists. A discussion of these marvelously 
complicated mechanisms would lead us far astray from the subject of interest. More information 
on virus ''life'' cycle can be found in Ref. \onlinecite{Bakerreview} and on translation and 
transcription (a process for obtaining the information-carrying RNA from the DNA molecule) in 
textbooks on cell biology and chemistry (see e.g. Ref. \onlinecite{textbookcell}). 

In 1956 Crick and Watson \cite{crickwat} proposed that the spherical protein coating (or capsid) probably has a 
platonic polyhedral symmetry i.e. that it is built of identical proteins assembled in 
a polyhedral shell. Caspar and Klug \cite{casparklug} developed this notion further by noting that 
the quantity of information contained in 
the viral genome is quite small, so that only one or perhaps two to three different proteins 
can be produced from it. They considered different polyhedral shells made of identical proteins 
and deduced that icosadeltahedral ordering provides a structure in which all 
of the proteins are in surroundings that are to a best approximation equal of all the 
choices considered. They called this the principle of quasi-equivalence. The geometry behind the 
principle of quasi-equivalence is the one already discussed in the cases of icosadeltahderal geodesic domes 
and fullerenes.

\begin{figure*}[ht]
\centerline{
\epsfig {file=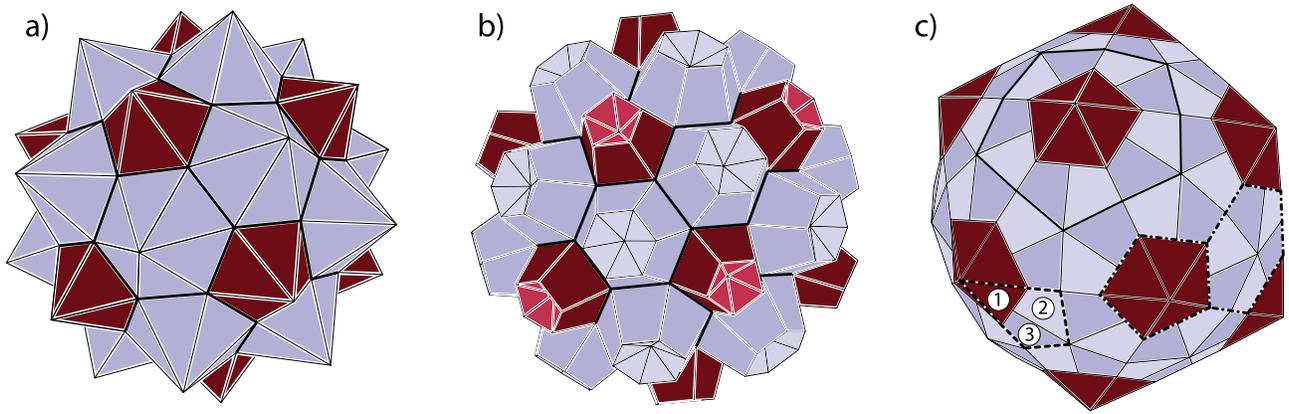,width=17cm}
}
\caption{Panels a) and b) represent polyhedral models of $T=3$ viruses. These polyhedra can be 
termed as omnicapped truncated icosahedra or omnicapped 
Buckminsterfullerenes. The ''pentamers'' are colored in a darker tone and borders between ''capsomers'' 
are represented by thicker lines. The polyhedron in panel b) is quite similar 
to turnip yellow mosaic virus. Panel c) represents a model $T=1$ (pT3) virus whose building block is a ''protein 
trimer'' outlined by dashed lines.} 
\label{fig:fig4}
\end{figure*}

In most of the ''spherical'' viruses, the proteins are grouped in clusters (capsomers) 
of five (pentamers) and six (hexamers). This is very often the case even if not all proteins are 
equal, i.e when capsid consists of several types of proteins.
In the assembled capsids, 
the twelve pentamers occupy the 
same spatial positions as carbon pentagons in fullerenes. The hexamers are equivalent to hexagonal 
carbon rings in fullerenes. Thus, viruses can be constructed by connecting each of 
the vertices in pentagonal and hexagonal rings of the fullerenes with the ring centers and interpreting the thus 
obtained divisions of the pentagons and hexagons as the dividing lines between the viral 
proteins. In fact, the fullerene vertices need not be connected to the centers of the rings but 
to points lying on approximate normals to the pentagonal and hexagonal faces and passing through 
centers of the faces. This corresponds to capping the pentagons and hexagons with pentagonal and 
hexagonal pyramids, respectively, but the capping can also be performed in many other ways. The thus obtained 
polyhedron may be called omnicapped 
fullerene [all ({\em omni}) of the fullerene faces capped by pyramids or some other polyhedra]. 
This procedure is illustrated panels a) and b) of Fig. \ref{fig:fig4}. The proteins are pieces of 
matter that have a certain equilibrium shape which is 
of course three-dimensional. Representation of protein capsomers by pyramids or any other polyhedron 
is thus approximate. Any three 
dimensional shape erected above the hexagon (pentagon) and having a six-fold (five-fold) 
symmetry with respect to rotations around the hexagon (pentagon) normal will serve as a representation 
of a viral hexamer (pentamer).

The classification of the symmetry of the capsid, however, does not depend on the shape of individual protein but 
only on the characteristics of the arrangement of all the proteins in the capsid (at least when all proteins 
are equal, see below). The symmetry of viruses is 
characterized in the same way as in the case of fullerenes: $m$ and $n$ integers are counted by 
''jumping'' {\em through the centers of the capsomers} and using the convention of turning left after 
the first $m$ jumps. If $m \ge n$, the virus is classified as a member of $T_{laevo}=m^2 + mn + n^2$ class 
(or simply $T$), and if otherwise, the virus is classified as a member of $T_{dextro}$ (or $T_d$) class. The 
virus-like polyhedra depicted in panels a) and b) of Fig. \ref{fig:fig3} both have $T=3$ symmetry, although the details of their 
shapes are quite different.

Total number of capsomers ($c$) in a T-class virus is obviously the same as the number points in the 
icosadeltahedral dome of T-symmetry, 
\begin{equation}
c = 10T + 2.
\label{eq:ncaps}
\end{equation}
Total number of proteins in a virus ($p$) is a sum of 60 proteins in 12 pentamers and $60(T-1)$ proteins in 
$10(T-1)$ hexamers. Alternatively (see Fig. \ref{fig:fig3}) one can deduce that $p$ for a virus 
of T-class should be the same as the number of faces in a dome of 3T-symmetry, i.e.
\begin{equation}
p = 60 T,
\label{eq:nprot}
\end{equation}
so that both approaches give the same answer.

There may occur problems in identifying the symmetry of the capsid when several proteins form a capsid, or when 
building blocks of the capsid are not pentamers and hexamers but trimers (clusters of three proteins). The 
problem is illustrated by the shape in panel c) of Fig. \ref{fig:fig4}. The building block of this shape is 
a protein trimer outlined by thick dashed lines. It consists of a darker triangular protein (denoted by 1) 
and two brighter kite-shaped proteins (denoted by 2 and 3). This structure could be identified as belonging to $T=1$ class, with only twelve 
''pentons'' [outlined by thick full lines in Fig. \ref{fig:fig4}c)] composed of five protein trimers 
(180 proteins in total). On the other hand, we could at least conceptually arrange the proteins in pentamers 
and hexamers as indicated by thick dash-dotted lines in Fig. \ref{fig:fig4}c). In this case, hexamers 
would contain three pairs of 2- and 3-proteins from three trimers, while pentamers would consist of 
five 1-proteins from five different trimers. This would implicate that the shape belongs to $T=3$ class 
of symmetry. Such viruses are called pseudo-T3 (or pT3) viruses. The problem with the identification could 
be resolved on physical grounds - if the binding energy between the proteins of the trimer is larger than 
between the proteins from different trimers, it makes sense to speak about the trimer as the basic building 
block and to identify the structure as belonging to $T=1$ class. However, the problem in the mathematical 
sense occurs when there are two $T$ numbers that can be divided without remainder (e.g. $T=9$ and $T=3$). In the most trivial 
case, every capsid with T-number $T_1$ could be interpreted as a $T=1$ capsid consisting of $T_1$-mers. In addition, 
every $(m,m)$ capsid could in principle be thought of as $(m,0)$ capsid made of trimers. The important question 
is again whether the conceptually obtained protein multimers make any sense as the strongly bonded elementary units. 

Caspar and Klug quasi-equivalence principle predicts that there are only twelve pentameric capsomers, all 
other being hexameric. In year 1982 it became clear that there are viruses (e.g. SV-40 virus from polyomavirus genus) 
composed {\em only} of pentamers, but still retaining the icosadeltahedral symmetry of their arrangement \cite{Rayment}. 
The concept of T-number is still valid in that case and the number of capsomers is still given by Eq. (\ref{eq:ncaps}),  
but the total number of proteins is no longer given by Eq. (\ref{eq:nprot}). For such viruses, the total number of 
proteins is
\begin{equation}
p = 5c = 5(10T + 2).
\end{equation}

Viruses, as fullerenes, should be considered as physical objects, i.e. their precise shape should be a result 
of interactions acting between proteins themselves and possibly between the proteins and viral DNA or RNA 
molecule. Shapes of viruses have recently been explored using a generic model of shells with icosadeltahedral 
symmetry \cite{Nelson,Bruinsma,Sibervir}. Application of physical methods to understand the energetics and 
assembly of viruses has been an area of very lively research in recent years (for a partial but quite 
readable glimpse of the field see Ref. \onlinecite{ZlotnickPNAS}).

This work has been supported by the Ministry of Science, Education, and Sports of Republic of Croatia 
through Project No. 035-0352828-2837, and by the National Foundation for Science, Higher Education, and 
Technological Development of the Republic of Croatia through Project No. 02.03./25.


\begin{thebibliography}{99}
\bibitem{primjedba1} One should be somewhat careful when using the terms pentagon (hexagon) and 
pentagonal (hexagonal) carbon rings as synonyms. All of the carbon atoms in a pentagonal or 
hexagonal ring do not necessarily lie in a plane, so that a pentagonal (hexagonal) carbon ring could in principle 
be built up of three (four) triangular pieces not lying in a plane. The angles between these pieces 
are in general very small, so that for all 
practical purposes, the carbon atoms in rings may be considered as belonging to the same plane.
\end{thebibliography}
\end{document}